\documentclass[fleqn,10pt]{wlscirep}
\usepackage[utf8]{inputenc}
\usepackage[T1]{fontenc}
\usepackage{braket}

\usepackage{changes}
\definechangesauthor[name={S.~M.}, color={blue}]{sm}

\title{Computing with two quantum reservoirs connected via optimized two-qubit non-selective measurements}

\author[1,*]{Stephen Vintskevich}
\author[1,2,*]{Dmitry Grigoriev}

\affil[1,*]{Moscow Institute of Physics and Technology,
Institutskii Per. 9, Dolgoprudny, Moscow Region 141700, Russia}
\affil[2,*]{Bitronics NeuroLab,  Likhachevskiy Proyezd 4, Dolgoprudny, Moscow Region 141701,  Russia}
\affil[*]{vintskevich@phystech.edu}

\begin{abstract}
Currently, quantum reservoir computing is one of the most promising and experimentally accessible techniques for hybrid, quantum-classical machine learning. However, its applications are limited due to practical restrictions on the size of quantum systems and the influence of noise. Here we propose a novel approach to connect two quantum reservoirs in a network to overcome these issues and enhance their computing performance. To transfer information between quantum reservoirs, we perform optimized two-qubit non-selective measurements. We suggest a general heuristic optimization strategy based on tensor network language and matrix representation of two-qubit quantum channels specified for quantum reservoir computing. In addition, we introduce a single qubit purification channel and its optimization for further enhancement of quantum reservoir computing. We also demonstrate that the optimized channels applied to a seven-qubit network can efficiently transfer information between its parts with the resulting performance comparable to a network up to twenty-five qubits connected via a classical information link.
\end{abstract}
\begin{document}

\flushbottom
\maketitle
\thispagestyle{empty}

\section*{Introduction}

In the seminal paper \cite{FujiiNakajima2017}, the authors proposed utilizing complex dynamics of multipartite quantum systems - quantum reservoirs (QR) - to perform learning and computational tasks with time-dependent data. 
This approach was named Quantum Reservoir Computing (QRC) as a counterpart to classical Reservoir Computing (RC) - a powerful framework that stems from recurrent neural network theory (see work \cite{RCBOOK2021} and references therein).
The detailed QRC framework is depicted in Fig. (\ref{fig:QRC_standard}).
Formally, the components of QRC can be divided into four main parts. 
First, valuable information is encoded in a state of some quantum system - denoted by red arrow and blocks marked by symbol $\Phi_{e}$ Fig.(\ref{fig:QRC_standard}) - (a).
It is then processed and transformed via dynamics of a QR - blocks with symbol marked by $\mathcal{U}_{R}$ in Fig.(\ref{fig:QRC_standard}) - (a,c) represents unitary evolution. The third is performing measurements of some quantum observable to gather processed encoded information (also named output signal). Finally, these output signals - measurement results - are used to train a simple linear model. To implement QRC in practice, one usually operates with a huge ensemble of identically prepared quantum systems (see Fig. \ref{fig:QRC_standard} - (a)) to effectively perform measurements using multiple copies of a system.
As a physical realization of an ensemble, one can utilize nuclear magnetic resonance (NMR) spin systems\cite{NMRCory2000,NMR2006,NMRRyan_2009, Jones2011}, semiconducting quantum dots \cite{schutz2016quantum}, superconducting qubits \cite{Quantum_simulation2014} and bosonic networks\cite{Mujal2021,Nokkala2021}. The great potential and scope of QRC have been recently demonstrated not only for processing classical temporal signals but also for quantum state preparation \cite{GhoshPRL} and reconstruction\cite{Ghosh2020}, training of an entanglement witness operator\cite{Ghosh2019}, and quantum tomography \cite{tran2021learning}. 

In contrast to other developing methods, such as quantum neural networks \cite{Killoran2019}, quantum circuit learning \cite{Mitarai2018,McClean2018}, and quantum machine learning \cite{Maria2019,Schuld2014}, intrinsic parameters of QRs and their dynamics are usually assumed to be fixed and non-tunable.
However, it is important to emphasize recently were presented methods of enhancements and adjustments, such as boosting QR's computational power through spatial multiplexing technique  \cite{Nakajima2019APL} and analyzing QR's intrinsic dynamics for better performance \cite{Kutvonen2020}. The latter is similar to hyperparameter tuning in neural networks training.
A recent approach that addresses the scalability and efficiency of QRC is called High Order Reservoir Computing (HQRC\cite{tran2020higher}), which is worth mentioning within the discussed above context.
In this paper, the authors proposed adjustments and enhancements of QRC by connecting multiple small QRs with five qubits each via a classical information link.
Such an approach is promising for realistic experimental implementations of QRC based on modern quantum technology \cite{lu2017enhancing}. Additionally, QRC may be interpreted as the 'inverse' to recently proposed methods that apply machine learning techniques to learn non-Markovian quantum dynamics of a qubit system in a complex environment \cite{LVGF,luchnikov2021probing}. 

In this work, we concentrate on further developing the idea of constructing small networks of multiple QRs. We propose a novel approach based on an action of a quantum channel that creates correlation and information transfer between two QRs and acts on just two-qubits. A quantum channel is a completely positive, trace-preserving linear map (CPTP) on a set of trace class operators that acts on some Hilbert space \cite{Heinosaari2009}. For simplicity, we named this channel a {\it{bridge}}. As a particular implementation of the bridge channel, we perform two-qubit nonselective projective measurements (measurements without reading the outcome) in a given two-qubit basis. 
As it was pointed out in work\cite{MEAS2013}, nonselective measurements can be used to transfer information between quantum systems and, as we demonstrated, can be used to improve QRC. 
One of the most vivid properties of multipartite quantum systems - entanglement must be presented on a particular measurement basis to transfer information between different QRs. Interestingly, it was proven \cite{NONSELECT} that nonselective projective measurement might even entangle two initially disentangled qubit subsystems, but the result substantially depends on a chosen basis. 
It was also pointed out \cite{NONSELECT} that projective nonselective measurements along a maximally entangling basis are useless for any entanglement gain between arbitrary two-qubit subsystems. However, even if the measurement two-qubit basis is an entangled basis, one does not guarantee good QRC performance.
To find an optimal configuration of a bridge channel, we propose a heuristic optimization method. 
The proposed optimization method is based on the language of tensor networks, matrix representation of quantum channels, and simple ideas of information transfer based on linear algebra.
We utilize a general two-qubit unitary operator ansatz to parameterize the bridge channel, which depends on 15 parameters as a result.
To numerically optimize the channel's parameters, we employ commonly used optimization algorithms in machine learning, such as Adam\cite{kingma2014adam}. 
Our optimization method does not depend on a particular QRC task or encoded information in QRs and can be considered as an adjustment of QRs' dynamics.
We also added and optimize auxiliary purification channels acting on bridge's qubits, which proved are being useful for enhancement of QRC performance. 
We utilize the same ansatz that we are using for the bridge optimization but with a slightly different optimization heuristic. Details are presented in the {\textbf{Methods}} section.

We numerically simulate the dynamics of two QRs (denoted by QR-$A$ and QR-$B$) for performance analysis of QRC assisted by optimized bridge and purification channels. 
For simplicity, we assume that all valuable information is encoded in a single qubit of QR-${A}$ named encoding qubit. 
We separately analyze two cases: extracting output signal only from  QR-${B}$ subsystem and extracting output signal from both QR-${A}$ and QR-${B}$.
Note, QR-${A}$ and QR-${B}$ are connected only via the bridge channel. 
Based on these simulations we analyze standard metrics for short-term memory tasks (STM)\cite{Kutvonen2020}, which gives us an understanding of the memory capability of QRs, one of the essential characteristics for successful QRC\cite{FujiiNakajima2017}.
Finally, we discuss possible new avenues in research and comparison with other setups (e.g., HQRC \cite{tran2020higher}).

\section*{Results}
\subsection*{General framework.} 
\begin{figure}[ht!]
\centering
\includegraphics[width=17.5cm]{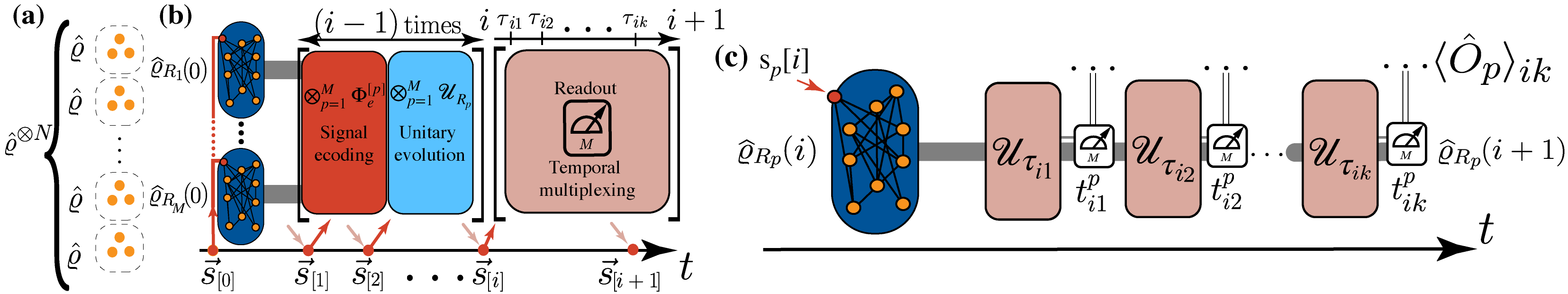}
\caption{The schematic of a general QRC framework. {\textbf{(a)}} - The idea of physical realization. In practice, one prepares identical physical systems (spins, atoms, etc.) using similar initial conditions. Signal encoding procedures and measurements are performed on each subsystem to get output signals  - ensemble mean values of a given observables. 
{\textbf{(b)}} - Dynamics of a QRC framework represented as a tensor network. An initial joint state of all QRs is factorized $\hat{\varrho}[i] = \otimes_{p =1}^{M}\hat{\varrho}_{R_{p}}[i], p = 1,\dots M$; for $i-1$ steps the dynamics of each QR is governed by the alternation of encoding-$\Phi^{[p]}_{e}[\bullet]$ and the unitary evolution-$\mathcal{U}_{R_{p}}$ channels followed by the output signal readout step, channels and states are connected via convolution of multi-indices (dark lines).
The input signal $\vec{s}_{i}$ (red arrows) at timestep $i$ can be mixed with QRs' output signals (beige arrows) at timestep $i-1$. {\textbf{(c)}} - readout step with temporal multiplexing. 
Time interval $[i, i+1],\ \forall i\in [1, i_{max}]$ is split into $K$ smaller sub-intervals $\tau_{ik}$ for each reservoir. For each timestep $t^{p}_{ik} = t^{p}_{i +(k/K)}$ one performs measurements to extract output signal $\langle\hat{O}_{p}\rangle_{ik}$.}
\label{fig:QRC_standard} 
\end{figure}

In this subsection, we provided an outline of a standard QRC framework. We considered a model of multiple reservoirs $R_{p}, p = 1,\dots M$ with similar structures, each consisting of $n_{p}$ qubits. A quantum state of a particular QR at the timestep $i$ is described by the positive Hermitian density operator $\hat{\varrho}_{p}[i]$ with the unit trace ${\rm{tr}}\left(\hat{\varrho}_{p}[i]\right) = 1, \forall i$.
In the case of multiple reservoirs, we assume that all reservoirs are completely uncorrelated at an initial time: $\hat{\varrho}[i] = \otimes_{p =1}^{M}\hat{\varrho}_{p}[i]$. The unitary evolution dynamics of each QR is governed by a fully connected Ising-type Hamiltonian $\hat{H}_{R_{p}} = \sum J^{[p]}_{ij}\hat{X}_{i}\hat{X}_{j} + h^{p}_{i}\hat{Z}_{i}$, where $\hat{X}$, $\hat{Z}$ are Pauli operators. 
We utilize tensor network notations (we recommend, e.g., works \cite{Ors2019,LVOF} and references therein) to describe the dynamics of QRs, which is shown in 
Fig.(\ref{fig:QRC_standard}).
More details regarding tensor notations and tensor networks are provided in {\textbf{Methods}}. In general QRC approach one operates with a classical discretized temporal signal $\vec{s}(t) \rightarrow \vec{s}_{[i]}$.
The encoding of signal in QRs is performed via a set of encoding quantum channels:
\begin{equation}\label{encoding}
    \{{\Phi}_{e}^{[p]}\}: {\Phi}_{e}^{[p]}[\hat{\varrho}] = \hat{\varrho}_{(\bullet_p)}(\vec{s}[i])\otimes{\rm{tr}}_{ (\bullet_p) }\left(\hat{\varrho}\right).
\end{equation}
Channels in Eq.(\ref{encoding}) map the initial state of the overall QR system into a tensor product of states corresponding to two subsystems: a state of a subsystem for encoding and a state of the rest of QR.
State of encoding subsystem (denoted as $\bullet_p$) is carrying valuable information about the signal, while ${\rm{tr}}_{ (\bullet_p)}(\hat{\varrho})$ is a state of other QR's subsystems. Symbol ${\rm{tr}}_{ (\bullet_p)}$ denotes partial trace operation. Thus the subsequent application of encoding and evolution channels maps this valuable information from the input signal to a $4^{N_{total}}$-dimensional vector in the linear space of trace class operators representing a quantum state of QR.
We denote $N_{total} = \sum_{p = 1}^{M}n_{p}$ to be the total number of qubits in the QR system.
We use a random binary signal as an input in the present work, which is the standard practice in QRC. In principle, one can mix signals' inputs at particular timesteps and the outputs from different QRs on a previous timestep. For instance, it was proposed to connect outputs from different reservoirs via a classical information channel   (HQRC)\cite{Nakajima2019APL,tran2020higher}.

To increase the performance of the QRs, one usually uses the temporal multiplexing technique \cite{FujiiNakajima2017} by performing $K$ additional measurements in between the subsequent timesteps $i$ and $i+1$ to get $\langle\hat{O}_{p}\rangle_{ik}$ (see also Fig. \ref{fig:QRC_standard}-\textbf{(b)}). We denoted $\langle\hat{O}_{p}\rangle_{i}$ as the output signal at the discrete timestep $i$. In the special case of just two reservoirs (QR-${A}$ and  QR-${B}$), one can measure the following observable: $\hat{O}(\{l_{a}, l_{b}\}) = (\otimes_{a=1}^{n_{A}}\hat{Z}_{a}^{l_{a}})\otimes(\otimes_{b=1}^{n_{B}}\hat{Z}_{b}^{l_{b}})$, where $\{l_{a}, l_{b}\} \in {0,1}$. 
Note that usually, a backaction of measurements on QR dynamics is not considered due to the possibility of performing simultaneous measurements of some observable for a big ensemble of identical reservoirs (Fig.\ref{fig:QRC_standard}).
It is assumed for each timestep $i$ that one has performed $i-1$ encoding steps which alternate with unitary evolution.
Then one begins to extract the output signal at timestep $i$ (Fig.\ref{fig:QRC_standard} - \textbf{c}). Thus, it is implied that one must repeat all the operations applied on QRs preceding timestep $t^{p}_{ik}$.  Note that for various $p$, the observables $\langle\hat{O}_{p}\rangle_{ik}$ and times $t^{p}_{ik}$ might be different.
In the present work, we consider the simplest form of the observable $\hat{O} \equiv  \hat{\mathbf{I}}_{AB/\bullet}\otimes\hat{Z}_{(\bullet)}$, where $\hat{Z}_{(\bullet)}$ is an operator that acts on a Hilbert space of an arbitrary readout qubit, and $\hat{\mathbf{I}}_{AB/\bullet}$ is unit operator.
Finally, the learning procedure with the QRC is straightforward.
In the supervised learning setup, we have a training sequence as a function of a timestep $i$ - $\mathbf{Y}[i]$. 
The objective is to minimize the mean square error (MSE): 
\begin{equation}\label{loss}
 \mathbf{W}_{*}= \underset{\mathbf{W}}{\rm{argmin}}|| \mathbf{Y}_{train} -  \mathbf{Y}_{mod.}(\mathbf{W})||_{2}; \ \mathbf{Y}_{mod.}(\mathbf{W}) = \langle \mathbf{O}\rangle_{out} \mathbf{W}^{T}.
\end{equation}
In Eq.(\ref{loss}) $\mathbf{Y}_{mod.}(\mathbf{W})$ is a linear model that weighs the set of output signals $\langle\mathbf{O}\rangle_{out}\equiv \{\langle\hat{O}_{p}\rangle_{ik}\}$, where $ik$ is reshaped in one multi-index, $||\bullet||_{2}$ is euclidean norm.
Dimensions (shapes) of the above matrices are as follows: $\mathbf{Y}_{train}$ - $ [T,L_{out}]$,$\langle\mathbf{O}\rangle_{out}$ - $[T,KN_{total}+1]$, and $\mathbf{W}$ - $[L_{out},KN_{total}+1]$. 
Thus, to train $\mathbf{W}$ one performs the Moore-Penrose pseudoinverse operation:
\begin{equation}\label{MOORE-PENROSE}
   \mathbf{W}_{*}^{T} = \left(\langle \mathbf{Z}\rangle^{T} \mathbf\langle {\mathbf{Z}}\rangle + \lambda \mathbf{I}\right)^{-1}\langle \mathbf{Z}\rangle^{T}\mathbf{Y}_{\rm{train}},
\end{equation}
where $\lambda$ is a regularization term. 
For our simulations, we set the regularization term's value $\lambda \sim 10^{-8}$. The value of the regularization term plays an important role in the overall performance of QRC. Moreover, simulations of QRC are sensitive even to small numerical variations of the output signals. One needs to tune the regularization term to avoid misinterpretations of the QRC simulation results and prevent the QRC algorithm from overfitting. In the following section we consider a new QRC setup assisted by the bridge and purification channels and analyze its performance.

\subsection*{A setup of two quantum reservoirs assisted by the bridge channel} 
\begin{figure}[ht!]
\centering\includegraphics[width=17.5cm]{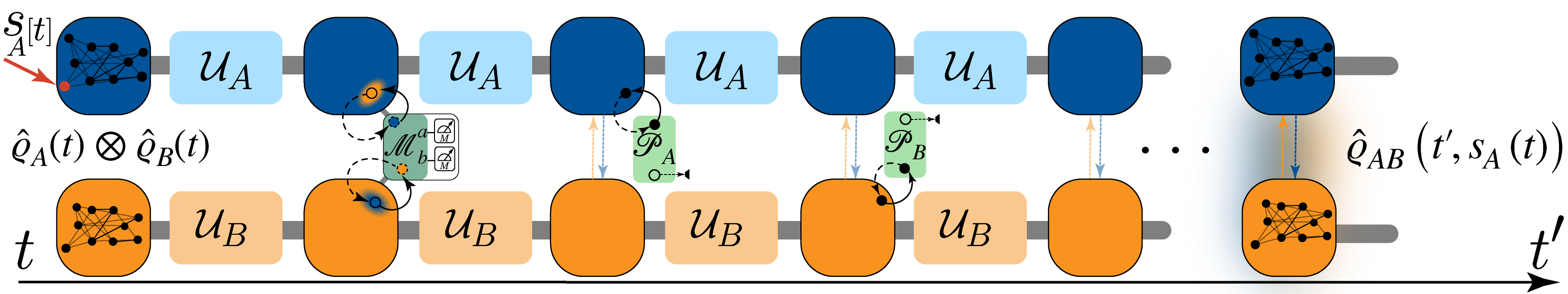}
\caption{The schematics of the QR's dynamics assisted by the bridge channel $\mathcal{M}_{b}^{a}$ which transmits of information between QR-$A$ and QR-$B$ represented as tensor network. The bridge channel is implemented as nonselective measurements on two chosen qubits (bridge qubits): $a$ from QR-$A$ and $b$ from QR-$B$. 
Auxiliary purification channels $\mathcal{P}_{A}$ and $\mathcal{P}_{B}$ subsequently act on the bridge qubits but separated by unitary evolution $\mathcal{U}_{A}$ and $\mathcal{U}_{B}$.
This alternation is necessary to spread the information inside each the QR's subsystems.
Only one qubit in the QR-$A$ is used for the signal encoding $S_{A}[t]$ (red arrow with $S_{A}[t]$). 
At an initial timestep $t$, it is assumed that the QR state is factorized.
The resulting correlated state (double arrows) of the whole QR at time $t'$  after all transformations is $\hat{\varrho}_{AB}(t', S_{A}[t])$.}
\label{fig:ESQR}
\end{figure}

\begin{figure}[ht!]
\centering
\includegraphics[width=17.6cm]{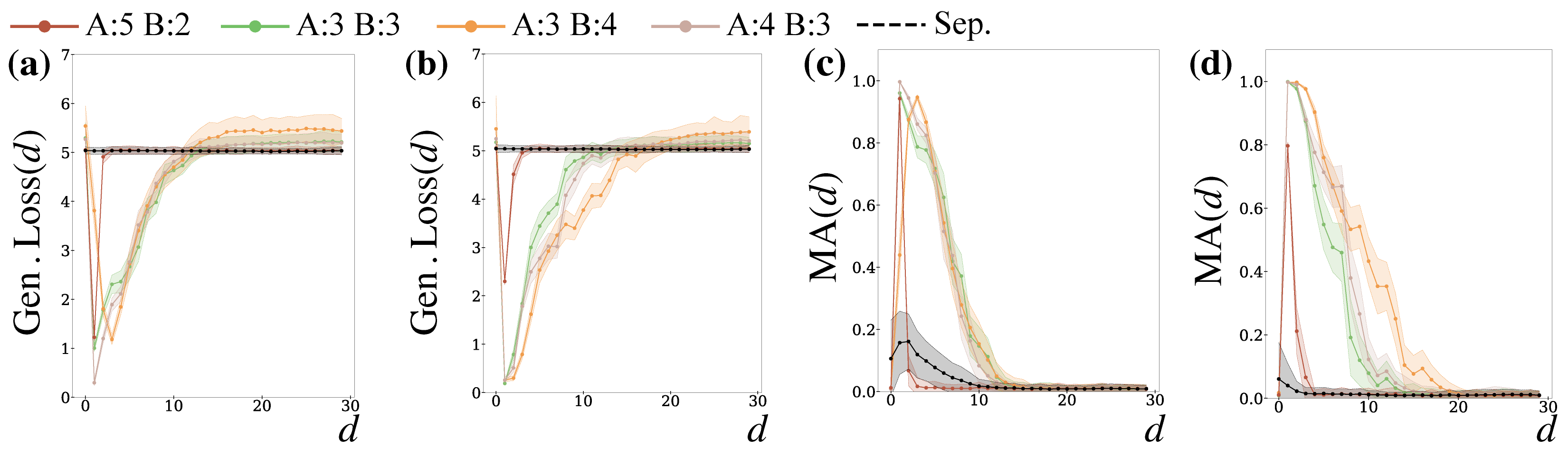}
\caption{Comparison of Generalization loss {\textbf{(a)}}, {\textbf{(b)}} and Memory accuracy {\textbf{(c)}},{\textbf{(d)}} with respect to the offset $d$ in the STM task without and with action of purification channels $\mathcal{P}$ respectively. The channel $\mathcal{P}$ is subsequently applied to each bridge qubit once per encoding step. Different lines correspond to the setups with different numbers of qubits in subsystems QR-$A$ and QR-$B$. We sampled 100 input signals and calculated average Generalization loss and Memory accuracy and its standard deviation (areas around average values) for all considered setups. 
The output signals are extracted only from the QR-$B$ to check the ability of the bridge channel to transfer the information. Note, the timestep of unitary QR dynamics is different for each setup and was found experimentally. We chose the timestep that minimizes the sum of Generalization losses for offsets from $d = 0$ to $d = 8$ in the STM task. The case (${\rm{Sep.}}$ - the dashed line) explicitly indicates that learning and generalization are impossible within the case of nonselective measurements in factorized bases.}
\label{CROSS_LEARNING}
\end{figure}

We consider a quantum system of two initially uncorrelated QRs (subsystems QR-$A$ and QR-$B$) at timestep $t$ to demonstrate our approach based on an action of the bridge channel.
We only use a single qubit of the QR-$A$ to encode the scalar signal in the QR system  (Fig.\ref{fig:ESQR}) via the standard encoding channel presented in the Eq.(\ref{encoding}).
The standard QRs dynamics, namely alternating action of encoding and unitary evolution channels, is now interrupted by the action of a bridge channel. The bridge channel is nonselective projective measurement $\mathcal{M}_{b}^{a}$ on two chosen qubits: $a$ from QR-$A$ and $b$ from QR-$B$ (see Fig.(\ref{fig:ESQR}) - {\textbf{(a)}}).
This channel $\mathcal{M}_{b}^{a}$ acts once between two subsequent encoding procedures. The action of the bridge channel on arbitrary QR's state $ \varrho_{AB}$ reads:
\begin{equation}\label{BRIDGE_DEF}
    \hat{\varrho}_{AB}' = \mathcal{M}_{b}^{a}[\hat{\varrho}_{AB}]  = \sum_{ij} \hat{\mathbf{I}}_{AB/a,b}\otimes\hat{\Pi}^{(a,b)}_{ij}(\vec{\theta}) \hat{\varrho}_{AB} \hat{\mathbf{I}}_{AB/a,b}\otimes\hat{\Pi}^{(a,b)}_{ij}(\vec{\theta}),
\end{equation}
where the operator $\hat{\Pi}^{(a,b)}_{ij}(\vec{\theta}) = \hat{U}(
\vec{\theta})\ket{i_{a}j_{b}}\bra{i_{a}j_{b}}\hat{U}^{\dagger}(\vec{\theta}), \ i_{a},j_{b} \in \{0,1\}$ is the projecive operator. 
Here we used standard notations $\{ \ket{i_{a}j_{b}}\},\ i_{a},j_{b} \in \{0,1\}$ for the two-qubit computational basis that in our case is transformed by the action of the unitary operator $\hat{U}(\vec{\theta})$. 
It allows one to parameterize a two-qubit basis used in our nonselective measurements. 
We denoted  $\vec{\theta}$ as the vector of parameters. 
Note that if the unitary operator $\hat{U}(\vec{\theta})$ is factorized as $\hat{U}(\vec{\theta}) = U_{a}(\vec{\theta})\otimes U_{b}(\vec{\theta})$, the resulting two-qubit measurements basis cannot be used to "connect" (create correlations between) the reservoirs QR-$A$ and QR-$B$. More details are provided in the \textbf{Methods} section.
We used the following ansatz \cite{Guan2014,KrausCirac2001} to represent arbitrary two-qubit operator $\hat{U}(\vec{\theta})$: 
\begin{equation}\label{ANSATZ}
\hat{U}(\vec{\theta}) = \hat{U}_{a}(\theta_{1\dots 3})\otimes \hat{U}_{b}(\theta_{4 \dots 6})e^{- i (\theta_{7} \hat{X} \otimes \hat{X}  + \theta_{8} \hat{Y} \otimes \hat{Y} + \theta_{9} \hat{Z}\otimes \hat{Z})}\hat{V}_{a}(\theta_{10 \dots 12})\otimes \hat{V}_{b}(\theta_{13 \dots 15}).
\end{equation}
The vector $\vec{\theta}$ in Eq.(\ref{ANSATZ}) contains fifteen parameters.
Unitary operators $\{\hat{U},\hat{V}\}_{a,b}$ are parameterized with three parameters each as standard qubit rotation operators along an arbitrary axis \cite{Nielsen2009}. 
We derived the particular $\vec{\theta}_{*}$ utilized in our simulations using a heuristic optimization strategy (see the \textbf{Methods} section).
It is important to demonstrate that QR's dynamics with the bridge channel can indeed transfer information from QR-$A$ to QR-$B$. As a demonstration we analyze the performance of setups where we extract output information only form the QR-$B$ with various numbers of qubits in QR-$A$ and QR-$B$ Fig.(\ref{CROSS_LEARNING}). 
We utilize the temporal multiplexing technique \cite{FujiiNakajima2017} in all further calculations, setting $K = 10$ as our multiplexing parameter. 

\begin{figure}[ht!]
\centering
\includegraphics[width=17.6 cm]{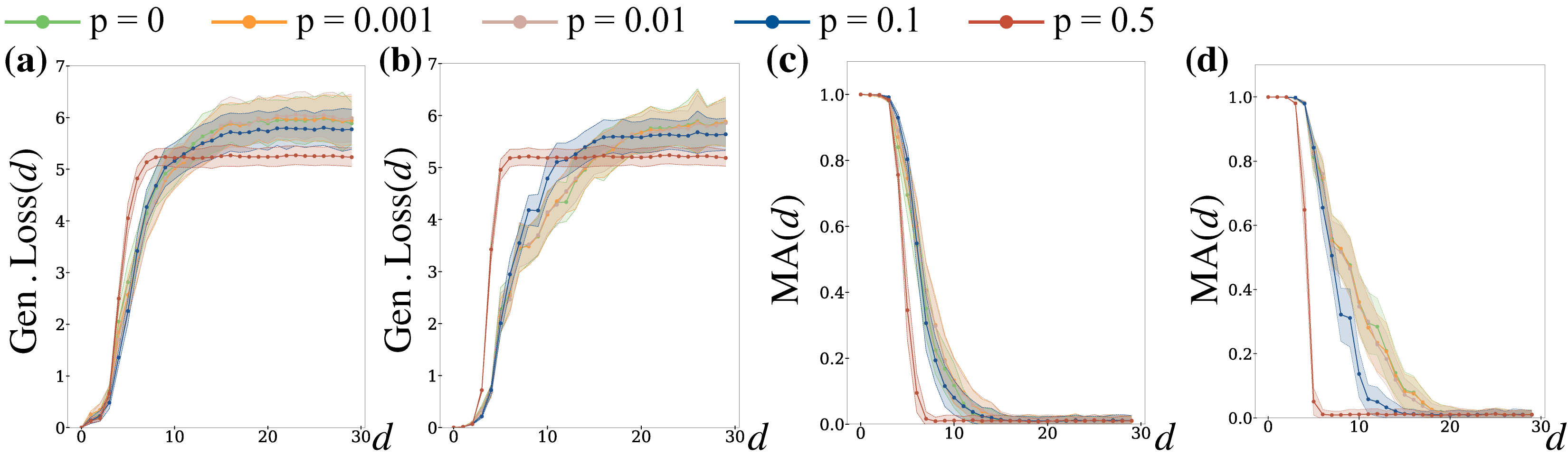}
\caption{Comparison of Generalization loss {\textbf{(a)}}, {\textbf{(b)}} and Memory accuracy {\textbf{(c)}},{\textbf{(d)}} without and with action of purification channel $\mathcal{P}$ respectively in the case of a setup with 3 qubits in QR-$A$ and 4 qubits in QR-$B$. The system is affected by the noise described by depolarization channel \eqref{Depolarization}; Different lines correspond to different depolarization probabilities ${\rm{p}} = 0,0.001,0.01,0.1,0.5$. Output signals are extracted from both QR-$A$ and QR-$B$ to check the QRC performance of the whole QR setup. Clearly, the QRC performance is stable against depolarization with probability up to ${\rm{p}} = 0.1$. In the most setups the action of purification channel can mitigate influence of depolarization.}
\label{ALL_P}
\end{figure}

In general one can extract the output signal from each QR's qubit. For extraction of output signals one performs multiple measurements of an observable over all ensemble for a chosen "extraction" qubit, including bridge's qubits $a$ and $b$ (see Fig. (\ref{fig:QRC_standard})). However, the action of the bridge channel severely affects the purity $P = {\rm{tr}}\hat{\varrho}_{\bullet}^2$ of any QR's subsystems states including states of $a$ and $b$, which might worsen the QRC performance. In addition, to simulate the influence of noise on QR's dynamics. We consider an action of the depolarization channel on each qubit (excluding qubits under action of a bridge channel) of the QR:
\begin{equation}\label{Depolarization}
\Phi_{dp.}({\rm{p}})[\bullet] = {\rm{p}}/2\hat{\mathbf{I}} + (1 - {\rm{p}})[\bullet],
\end{equation}
where ${\rm{p}} \in [0,1]$ - is the probability of depolarization for the chosen qubit. 
To circumvent this problem, we introduce an ancilla qubit system in a fiducial pure state $\ket{0}_{a}$. Interaction with an ancilla can be used to purify a particular single qubit subsystem \cite{Cirac1999}. 
In our simulations, we deliberately assume that one has access only to the bridge qubits and only one encoding qubit in the QR's subsystem $A$.
Thus, a purification channel acts on the bridge qubits $a$ or $b$ - Fig.(\ref{fig:ESQR}) - ($b_{1}$,$b_{2}$) and can be written as follows:
\begin{equation}\label{Purification}
\mathcal{P}_{\bullet}[\hat{\varrho}_{AB}] = {\rm{tr}}[\hat{\mathbf{I}}_{AB/\bullet}\otimes\hat{U}_{\bullet}\left(\gamma\right)\hat{\varrho}_{AB}\otimes \ket{0}_{a a} \bra{0}\hat{\mathbf{I}}_{AB/\bullet}\otimes\hat{U}_{\bullet}^{\dagger}\left(\gamma\right)].
\end{equation}
We optimize the purification channel using the same ansatz for the unitary operator $\hat{U}_{\bullet}\left(\gamma\right)$ that was used to optimize the bridge channel Eq.(\ref{ANSATZ}). However, the optimization heuristic for purification channel is different. It is based on the trade-off between purity $P$ of the particular qubit and the fidelity $F={\rm{tr}}(\sqrt{\sqrt{\hat{\varrho}^{out}}\hat{\varrho}^{in}\sqrt{\hat{\varrho}^{out}}})$ calculated for its state before and after purification. Based on the ansatz optimization result, we indeed achieved a significant QRC performance improvement for different setups with and without depolarization. Comparison  results of QRC performance with and without depolarization are presented in Fig.(\ref{CROSS_LEARNING}) and Fig. (\ref{ALL_P})).

Finally, to quantify the QRC performance of the proposed approaches, we focus on a commonly used QRC task - short-term memory task (STM)\cite{Kutvonen2020}.
We analyze this task using several metrics.
The first is memory accuracy ${\rm{MA}}(d)$ \cite{jaeger2002short} given by
\begin{equation}\label{MA}
{\rm{MA}}(d) ={\rm{cov}}^2(\mathbf{Y}_{mod.}, \mathbf{Y}_{test.}(d))/({\rm{var}}(\mathbf{Y}_{mod.}){\rm{var}}(\mathbf{Y}_{test.}(d))),
\end{equation}
where $Y_{mod}$ is the prediction of QRC, and $Y_{test.}(d) = s[i-d]$ is the input signal, shifted by $d$ steps. 
Symbols $\rm{var}$, $\rm{cov}$ represent standard variance and covariance. 
We take a random binary signal as an input signal to the STM task.
Secondly we quantify the Generalization loss which is euclidean length (the $L_2$ vector norm) of the prediction error: ${\rm{Gen.Loss}} = ||\mathbf{Y}_{mod.} - \mathbf{Y}_{test.}||_{2}$. 
%\begin{figure}[ht!]
%\centering
%\includegraphics[width=11.5cm]{NALL_LEARNING.png}
%\caption{The metrics and setup are identical to Fig.\ref{ALL_P} but purification channel is not applied. In the most setups the action of purification channel can mitigate influence of depolarization (see Fig. \ref{ALL_P}).}
%\label{ALL_NP}
%\end{figure}
\begin{figure}[ht!]
\centering
\includegraphics[width=17cm]{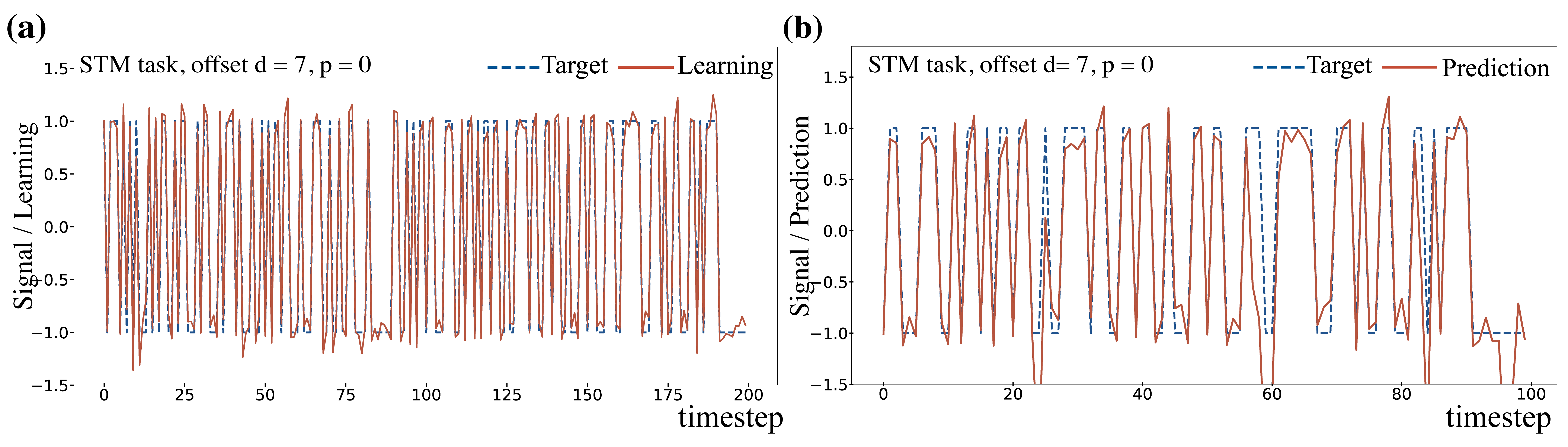}
\caption{Visualization of the STM task performance on a random binary input signal for a setup with 3 qubits in  QR-$A$ and 4 qubits in QR-$B$.
STM task offset $d = 7$ corresponds to a good memory accuracy (see Fig. \ref{ALL_P}). 
The left and right plots depict the performance of the algorithm on the training {\textbf{(a)}} and testing data {\textbf{(b)}} sets respectively.
Depolarization is not taken into account (${\rm{p}} = 0$). 
The purification channel is applied.}
\label{LENGEN}
\end{figure}

In this section, we demonstrated the usefulness of the auxiliary purification channels. We compared various setups with and without the action of purification channels in Fig.(\ref{CROSS_LEARNING}), Fig. (\ref{ALL_P}).
The Fig.(\ref{CROSS_LEARNING}) shows the STM task with the output signal extracted only from the subsystem $B$ for various number of qubits. This figure clearly validates that the bridge channel can indeed be used to transfer quantum information between quantum reservoirs. The Generalization loss in Fig.(\ref{CROSS_LEARNING}) - {\textbf{(a)},\textbf{(b)}} has a spike at $d = 0$. It requires some time to propagate from the encoding qubit in QR-$A$ to the readout qubits in QR-$B$. 
So most of the outputs from QR-$B$ at timestep $i$ have no information about the signal $s_i$. Results in Fig.(\ref{CROSS_LEARNING}) also shows that the action of purification channel $\mathcal{P}$ increases the performance in most setups. The figure (\ref{ALL_P}) focuses on affection of depolarization and emphasizes positive influence of purification channel on QRC performance. In some cases QRC performance increases by almost 50 \%. Finally, Fig. (\ref{LENGEN}) explicitly demonstrates the performance of one chosen setup analyzed in Fig. (\ref{ALL_P}) without action of depolarization noise (${\rm{p}} = 0$). The resulting generalization shows good performance on the STM task up to the offset $d=7$. Our analysis concludes that the bridge channel can transfer valuable information for QRC, and proposed setups are stable to noise, especially when the purification channel assists QR dynamics.
\section*{Summary}
It is clear that due to practical limitations caused by control difficulties and a magnitude of noise in quantum systems, the accessible quantum-enhanced feature space of QR is also limited. 
In practice, the construction of relatively small QRs is one of the most realistic solutions that increase the scalability of QRC.
In this paper, we suggested a novel way to connect QRs in a network. 
We proposed utilizing two-qubit nonselective measurements in two-qubit entangled basis to transfer information between QRs within the network. 
This channel, denoted as the bridge, allows one to make a network of reservoirs without any classical communication between them. 
The proposed approach excludes classical information processing for the QRC procedure up to the final linear optimization step. 

In addition, we introduced auxiliary single-qubit purification channels to further enhance QRC performance.
Our approach can be treated as an intermediate scheme in-between QCL and QRC. 
The presented numerical optimization algorithms for both the bridge and the purification channels are based on the application of tensor network language, and the analysis of matrix representation of quantum channels. 
Note that the optimization strategy is unconditional - it does not depend on a particular QRC task or its structure and does not require a lot of computational resources. The optimization strategy can also be extended further by considering various parametrizations of interactions between multiple QRs.
We demonstrated the performance of two QRs connected via the bridge channel by numerical simulations. 
We observed that the optimized bridge channel with just 7 qubits produces memory accuracy comparable with a classically connected QRs network up to 25 qubits. 
We also performed simulations to validate that the purification procedure enhances the QRC performance affected by noise.

Another possible performance enhancement could be achieved by optimizing the channel application sequence and considering more than two-qubit subsystems.
The particular parameters of the given QRC setup should be optimized for various physical QR realizations: photons, atoms, molecules, etc. 
The most important problem for further investigation is analyzing the performance and scalability of setups with more than two QRs, which is the next step of our research.
We believe that further development of a proposed technique and enhancements of optimization strategies of information transfer in QRs networks will drastically increase the computational power of setups with an arbitrary number of interconnected QRs. As a result, adjustable QR networks will expand QRC applications, including classical and quantum information processing tasks.

\section*{Methods}\label{METHODS}
This section provides details regarding the tensor network language and proposes the heuristic optimization strategies for the given ansatz used for parametrization of the bridge \eqref{BRIDGE_DEF} and purification channels \eqref{Purification}.

In our analysis we utilize the commonly used language of tensor networks\cite{Ors2019}. It has a simple graphical representation of quantum dynamics of multipartite quantum systems \cite{LVOF}.
We have demonstrated QRs dynamic and schematic of QRC using this language in Fig. \ref{fig:QRC_standard} and Fig. \ref{fig:ESQR}.
To describe a QR's quantum state one can consider a basis in the joint Hilbert space $\mathcal{H}_{total} = \otimes_{p=1}^{p = M}\mathcal{H}_{p}$: $\{\ket{i_{1}, \dots i_{M}}\}$. 
Thus, $\hat{\varrho} = \sum_{\{i_{1},\dots i_{p}\}}\varrho_{i_{1},\dots i_{M}i'_{1},\dots i'_{M}}\ket{i_{1},\dots i_{M}}\bra{i'_{1},\dots i'_{M}}$.
The matrix element $\varrho_{i_{1},\dots i_{M}i'_{1},\dots i'_{M}}$ - is a multi-index array which we call tensor. 
It has rank $2M$, but one can rearrange these indices by reshaping this tensor into a vector (rank one tensor) by assembling all indices in one big multi-index: $i_{1},\dots i_{M}i'_{1},\dots i'_{M} \longrightarrow I, I \in  \{0, 4^{N_{total}}\}$. Such representation is handy for describing quantum dynamics, operating, and addressing subsystems of the whole QR. One can always rearrange tensor indices to form groups corresponding to particular subsystems of interest. In this case an arbitrary quantum channel acts on quantum state $\hat{\varrho}' = \Phi[\hat{\varrho}]$ can be represented as matrix (rank - 2 tensor): $\varrho_{I'} = \Phi_{I'I}\varrho_{I}$. Here, we utilized the Einstein index summation rule: each index can appear at most twice in any term. Repeated indices are implicitly summed over. Note, one can use alternative basis to represent a multipartite quantum state ${\hat{\varrho}}$ as a vector. For instance one considers 
a linear space of trace class operators \cite{Heinosaari2009} acting on Hilbert space $\mathcal{H}_{total}$. 
$\mathcal{T}\left(\mathcal{H}_{total}\right)$ with some operator basis $\{ \hat{E}_{j}\},j = 0,\dots 4^{N_{total}}$. Thus, an arbitrary quantum state has the following representation: $\hat{\varrho} = \sum_{j=0}^{4^{N_{total
}}}\varrho_{j}\hat{E}_{j}$ with real-valued elements $\varrho_{j}$. Alternatively, one can also employ Choi-Jamiolkowski isomorphism to represent quantum channels and their action on quantum states \cite{Heinosaari2009}. Our analysis and derivations operate with both computational basis for multi-qubit systems and multi-qubit Pauli operator basis. In the Fig. \ref{fig:QRC_standard} - {\textbf{(b)}} we represent state of each reservoir $R_{p}$ as a tensor with one multi-index and whole QR system has $M$ multi-indeces whereas quantum channels have $2M$ indices. Consecutive convolution of indices reflects QRs dynamics - change of $\hat{\varrho}(t)$ in time.

In the Eq.(\ref{BRIDGE_DEF}) of the main text we proposed the bridge channel - a way to use nonselective measurements to transfer information between QR-$A$ and QR-$B$. Additionally, to improve the QRC performance, we utilize a purification channel Eq.(\ref{Purification}). 
We use a general ansatz Eq. (\ref{ANSATZ}) to parametrize both quantum channels. 
Let us start by considering the bridge channel. 
Firstly, it is easy to show explicitly that the "presence" of entanglement is necessary for information transfer via the bridge channel. 
We assumed that information about a signal is encoded in QR-$A$ at an arbitrary time $t$. 
For simplicity, we implied that no information about a signal was presented in the QR system before the time $t$. 
Thus, an arbitrary initial state of the QR after the encoding can be written as follows: $\hat{\varrho}_{AB}(t) = \sum_{k}q_{k}\hat{\rho}_{A/a,a}^{[k]}(s,t)\otimes\hat{\rho}_{b,B/b}^{[k]}(t)$ where state  normalization is assumed: $\sum_{k}q_{k} = 1$. Coefficients $q_{k}$ do not depend on an encoded signal, and ${\rm{tr}}_{A}{\rm{tr}}_{B}(\hat{\varrho}_{AB}(t)) = {\rm{tr}}_{B}{\rm{tr}}_{A}(\hat{\varrho}_{AB}(t)) = 1$. If we use coefficients $\theta_{7},\dots \theta_{9} \equiv 0$ in Eq.(\ref{ANSATZ}) and substitute it into Eq.(\ref{BRIDGE_DEF}), the resulting state after the action of the bridge channel $\hat{\varrho}_{AB}^{'}(t) = \mathcal{M}_{b}^{a}[\hat{\varrho}_{AB}(t)]$ has the following form:
\begin{eqnarray}\label{sep_nonselective}
\hat{\varrho}_{AB}^{'}(t) = \sum_{i_{a},j_{b},k}q_{k}\left(\bra{i_{a}}\hat{Q}_{a}\hat{\rho}_{A/a,a}^{[k]}(s,t)\hat{Q}_{a}^{\dagger}\ket{i_{a}}\right)\ket{i_{a}}\bra{i_{a}}\otimes\left(\bra{j_{b}}\hat{Q}_{b}\hat{\rho}_{b,B/b}^{[k]}\hat{Q}_{b}^{\dagger}\ket{j_{b}}\right)\ket{j_{b}}\bra{j_{b}} =\sum_{k}q_{k}\hat{\rho}^{'[k]}_{A}(s,t)\otimes\hat{\rho}^{'[k]}_{B}(t),
\end{eqnarray}
where $\hat{Q}_{a,b} =\hat{U}_{a,b}\otimes\hat{V}_{a,b} $ are the unitary operators that act on subsystems $a$ or $b$.
Thus, for any observable $\hat{\mathbf{I}}_{A}\otimes\hat{O}_{B}$, the resulting mean value $\braket{\hat{\mathbf{I}}_{A}\otimes\hat{O}_{B}}$ does not carry any information about the encoded signal:
\begin{eqnarray}
\braket{\hat{\mathbf{I}}_{A}\otimes\hat{O}_{B}} = \sum_{k}q_{k}{\rm{tr}}_{A}(\hat{\rho}_{A}^{'[k]}(s,t)){\rm{tr}}_{B}(\hat{O}_{B}\hat{\rho}_{B}^{'[k]}(t)) = \sum_{k}q_{k}{\rm{tr}}_{B}(\hat{O}_{B}\hat{\rho}_{B}^{'[k]}(t)).
\end{eqnarray}
As a result, nonselective measurements parametrized with $\theta_{7},\dots \theta_{9} \equiv 0$  cannot lead to information transfer between the QR's subsystems.
One can use the channel matrix representation to gain more insights about the structure of a bridge channel. Let us fix an orthogonal operator basis set in space $\mathcal{T}\left(\mathcal{H}_{total}\right)$: $\{\hat{\sigma}_{a}\otimes\hat{\sigma}_{b}\otimes \hat{e}_{\xi}\}$.
Operators $\{\sigma_{a,b}; \ a,b \in \{0,\dots 3\}$ are standard Pauli operators $\{\hat{\mathbf{I}}, \hat{X}, \hat{Y}, \hat{Z}\}$ acting on qubits $a$ and $b$. 
Basis operators $\hat{e}_{\xi},\xi_{0,\dots 4^{\left(N_{total}-2\right)}-1}$ act on the subsystem QR-$AB/ab$. 
We explicitly distinguished basis operators corresponding to the subsystem of bridge qubits. 
Thus, we can also represent an arbitrary QR state $\hat{\varrho}_{AB}(t)$ as follows:
\begin{equation}\label{state_vector}
\hat{\varrho}_{AB}(t) = \sum_{ab,\xi} r_{ab,\xi}(t) (\hat{\sigma}_{a}\otimes\hat{\sigma}_{b})\otimes \hat{e}_{\xi}; \ r_{ab,\xi}(t) = {\rm{tr}}(\hat{\varrho}_{AB}(t) \frac{(\hat{\sigma}_{a}\otimes\hat{\sigma}_{b})}{4}\otimes \frac{\hat{e}_{\xi}}{2^{N_{total}-2}})
\end{equation}
The action of an arbitrary two-qubit channel $\mathbf{I}_{AB/ab}\otimes\mathcal{Y}_{b}^{a}[\hat{\varrho}_{AB}(t)]$ can be represented by:
\begin{eqnarray}\label{chennel_representation}
\hat{\varrho}_{AB}^{'}(t) = \mathbf{I}_{AB/ab}\otimes\mathcal{Y}_{b}^{a}[\varrho_{AB}(t)] = \sum_{ab,\xi}r_{ab,\xi}(t)\mathcal{Y}_{a}^{b}[\hat{\sigma}_{a}\otimes\hat{\sigma}_{b}]\otimes \hat{e}_{\xi} =  \sum_{ab,\xi}(\sum_{a'b',\xi'}\Upsilon_{ab,a'b'}\delta_{\xi,\xi'}r_{a'b',\xi'}(t))\left(\hat{\sigma}_{a}\otimes\hat{\sigma}_{b}\right)\otimes \hat{e}_{\xi},
\end{eqnarray}
where $\Upsilon_{ab,a'b'}$ is a matrix representation of a two-qubit channel $\mathcal{Y}_{b}^{a}[\bullet_{a,b}]$ and is given by: $\Upsilon_{ab,a'b'} = \frac{1}{4}{\rm{tr}}\left(\hat{\sigma}_{a}\otimes\hat{\sigma}_{b}\mathcal{Y}_{b}^{a}[\hat{\sigma}_{a'}\otimes\hat{\sigma}_{b'}]\right)$. 

\begin{figure}[ht!]
\centering
\includegraphics[width=16cm]{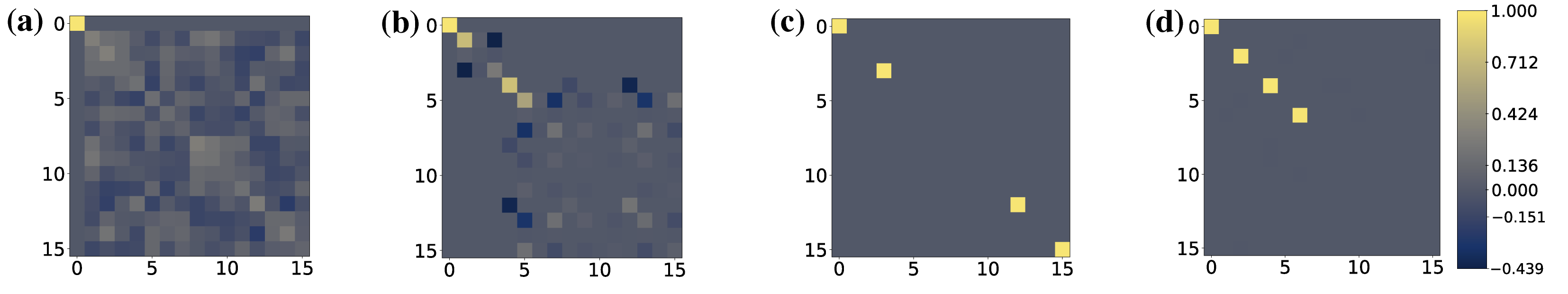}
\caption{
This figure presents the matrix plot of the bridge channel's matrix representation $M_{ab,a'b'}$.
{\textbf{(a)}} are the optimization results obtain via the strategy described in Eq.(\ref{Bridge_opt}).
The corresponding matrix $M_{ab,a'b'} (\theta_{*})$ is the most "diverse" and provides the best QRC performance among the considered cases.
{\textbf{(b)}} is the matrix plot in the case of nonselective measurements along a random factorized basis with ($\theta_{7} = \theta_{8} = \theta_{9}\equiv 0$).
{\textbf{(c)}} is the matrix plot in the case of nonselective measurements along a standard computational basis; this result corresponds to minimal $||M_{ab,a'b'}^{SEP}||_{1} = 4$.
{\textbf{(d)}} presents the results of the "anti" optimization strategy where we minimize the $1$-norm of the matrix $||M_{ab,a'b'}||_{1}$. 
We use the same values as in {\textbf{(c)}}: $\theta_{7*} = \theta_{8*} = \theta_{9*} \approx 0$ and minimal norm $||M_{ab,a'b'}||_{1} \approx 4$. 
We verified the optimization results {\textbf{(c)}} and {\textbf{(d)}} for various initial conditions $\vec{\theta}$.}
\label{Bridge_opt}
\end{figure}

As we mentioned earlier, both the bridge and the purification channels depend on the same ansatz Eq.(\ref{ANSATZ}), which implies parametrization of a particular channel and its matrix representation on parameters $\theta$: $\mathcal{Y}_{b}^{a}(\theta) \rightarrow \Upsilon_{ab,a'b'}(\theta)$. 
To optimize the bridge channel $\mathcal{M}_{b}^{a}\left(\theta\right)$ with the corresponding matrix representation $M_{ab,a'b'}\left(\theta\right)$ we use the following heuristic observation. 
In the case of nonselective measurements in factorized bases (no entanglement presence in ansatz Eq.(\ref{ANSATZ}), Eq.(\ref{sep_nonselective})) a $1$ - norm of the $M_{ab,a'b'}\left(\theta\right)$, in general, is less than it is in the case of measurements along entangled bases $||M_{ab,a'b'}^{SEP.}||_{1}\leq ||M_{ab,a'b'}^{ENT.}||_{1}$. 
The $1$-norm, by definition, is the maximum absolute column sum of a matrix. 
From Fig.(\ref{Bridge_opt}), it is clear why the $1$-norm is a good candidate.
Let us take a look at an arbitrary factorized two-qubit state $\hat{\varrho}_{ab} = \hat{\varrho}_{a
}\otimes\hat{\varrho}_{b} = \sum_{ab}r_{a}r_{b}\left(\hat{\sigma}_{a}\otimes\hat{\sigma}_{b}\right)$. 
The corresponding coefficients $r_{a}r_{b}$ in a matrix representation are also factorized.
Let us assume that $r_{a}$ is some function of the signal $s$ that carries some "valuable" information: $r_{a}\equiv r_{a}(s)$. Consequently, if one would like to transfer "valuable" information about the signal from $a$ to $b$, one needs to "mix" the corresponding coefficients and assemble a new non-factorized matrix $r_{ab}$. 
The matrix $r_{ab}$ also tends to have the maximum number of non-zero elements that nontrivially depend on the signal.
The diversity of transformations of valuable information inside a reservoir system is an essential feature of RC, both classical and quantum. 
Indeed, it means that the matrix representation of a bridge channel has to be as "diverse" as possible.
This observation can be directly verified via a plot of the matrix $M_{ab,a'b'}\left(\theta\right)$ for various cases (see Fig.\ref{Bridge_opt}). 
It is clear that the bridge channel's matrix in the case of nonselective measurements along the entangled basis in Fig.(\ref{Bridge_opt})-{\textbf{(a)}} is the most diverse matrix.
We also aim to find such matrix $M_{ab,a'b'}\left(\theta_{*}\right)$ to contain as many nonzero values as possible. 

\begin{figure}[ht!]
\centering\includegraphics[width=17.5cm]{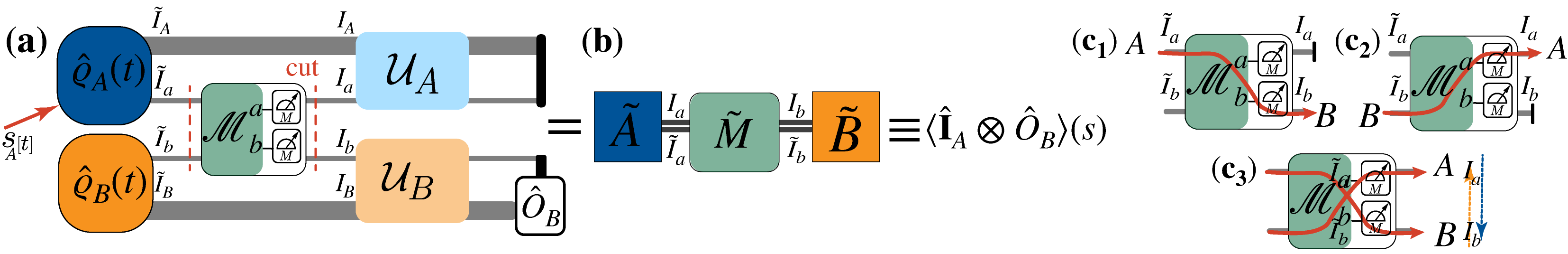}
\caption{ {\textbf{(a)}} The tensor network  describing the transfer of valuable information from  QR-$A$ to QR-$B$ via action of the bridge channel and extraction of processed information via measurements of some arbitrary  observable $\hat{O}_{B}$ in QR-$B$. The black bold line at the end of a tensor corresponds to full tensor convolution $\#$, as it was described after \eqref{tens_conv}.
{\textbf{(b)}} - rearranged indices in the network to explicitly separate bridge tensor ${\tilde{M}}_{[I_{a}\tilde{I}_{a}],[I_{b}\tilde{I}_{b}]}$ for further optimization of information transfer. Sub-figure $(c_{1})$: the schematic of information flow from QR-$A$ to QR-$B$; $(c_{2})$ - schematic of information flow from QR-$B$ to QR-$A$ and $(c_{3})$ - corresponds to exchange of information between both QRs, which is necessary for the effective QRC.}
\label{fig:tens_opt}
\end{figure}

The above heuristic optimization idea can be further strengthened by analyzing QR's dynamics using tensor network graphical representation. For this purpose, we consider a standard computational multi-qubit basis $\{\ket{i_{{1_{A}}}, \dots i_{n_{B}}}\}$. We explicitly separate indices corresponding to bridge's qubits subsystems ($\tilde{I}_{a},\tilde{I}_{b},I_{a},I_{b}$). Let us assume that arbitrary states of QR-$A$ and QR-$B$ are factorized at initial time moment $t$. Let us also consider an arbitrary observable $\hat{\mathbf{I}}_{A}\otimes\hat{O}_{B}$ acting on subsystem of QR-$B$. The resulting tensor network is depicted on Fig. (\ref{fig:tens_opt})-{\textbf{(a)}}. Based on the prescribed index notations, the mean value of the observable can be rewritten as a tensor convolution. (Einstein summation rule is assumed), see also Fig. (\ref{fig:tens_opt})-{\textbf{(b)}}:
\begin{eqnarray}\label{tens_conv}
\langle\hat{\mathbf{I}}_{A}\otimes\hat{O}_{B}\rangle(s) =  \tilde{A}_{[I_{a}\tilde{I}_{a}]}\tilde{M}_{[I_{a}\tilde{I}_{a}],[I_{b}\tilde{I}_{b}]}\tilde{B}_{[I_{b}\tilde{I}_{b}]}, \  \tilde{A}_{[I_{a}\tilde{I}_{a}]} = {\mathcal{U}_{A}}_{[\#],[\tilde{I}_{A}I_{a}]}\varrho_{[\tilde{I}_{A}\tilde{I}_{a}]}(s), \  \tilde{B}_{[I_{b}\tilde{I}_{b}]} = O_{J_{b}}{\mathcal{U}_{B}}_{[{J_{b}},\#],[\tilde{I}_{B}I_{b}]}\varrho_{[\tilde{I}_{B}\tilde{I}_{b}]} 
\end{eqnarray}
where we reshaped bridge tensor to the matrix $\tilde{M}_{[I_{a}\tilde{I}_{a}],[I_{b}\tilde{I}_{b}]}$ and combined multi-indeces in all tensors to emphasize role of $\tilde{M}_{[I_{a}\tilde{I}_{a}],[I_{b}\tilde{I}_{b}]}$ in information transfer. The symbol $\#$ - indicates full convolution of a given tensor along some set of indices, which reduce rank of a tensor over $2M$, with natural number $M$.
Thus, it is necessary that matrix $\tilde{M}_{[I_{a}\tilde{I}_{a}],[I_{b}\tilde{I}_{b}]}$ has rank more then 1 to transfer information from QR-$A$ and QR-$B$ and vice versa. From (\ref{fig:tens_opt})-{\textbf{(b)}} and \eqref{tens_conv}, it is also clear that the properties of information transfer do not depend on the inner structure of tensors $\tilde{A}$ and $\tilde{B}$ and their unitary transformations, including the action of an observable operator. It corresponds to simple criteria of information transfer form QR-$A$ to QR-$B$ by assuming  $\frac{\partial}{\partial s}\langle\hat{\mathbf{I}}_{A}\otimes\hat{O}_{B}\rangle(s) \neq 0$. On the other hand, it is clear that $\langle\hat{\mathbf{I}}_{A}\otimes\hat{O}_{B}\rangle (s) = {\rm{tr}}_{B}({\rm{tr}}_{A}(\hat{\varrho}_{AB}'(s))\hat{\mathbf{I}}_{A}\otimes\hat{O}_{B}) = {\rm{tr}}_{B}(\hat{\varrho'}_{B}(s)\hat{O}_{B})$. Consequently, based on \eqref{tens_conv} and Fig.(\ref{fig:tens_opt}) one needs to maximize 1-norm of rows of reduced tensor ${\tilde{M}}_{[\#\tilde{I}_{a}],[I_{b}\tilde{I}_{b}]}$ reshaped in matrix, owing to linear structure of tensor transformations to maximize information transfer from QR-$A$ and QR-$B$ (\ref{fig:tens_opt})-{${\rm{(c_{1})}}$}). 
The same idea works if we want to maximize information transfer form QR-$B$ to QR-$A$ (see Fig.(\ref{fig:tens_opt})-{${\rm{(c_{2})}}$}), but we need to maximize 1-norm of columns of corresponding reduced reduced tensor ${\tilde{M}}_{[I_{a}\tilde{I}_{a}],[\#\tilde{I}_{b}]}$ again reshaped in matrix. However for effective performance of QRC we need to exchange information between both QRs Fig.(\ref{fig:tens_opt})-{$\rm{(c_{3})}$}). This observation implies maximization of 1-norm of full tensor ${\tilde{M}}_{[I_{a}\tilde{I}_{a}],[I_{b}\tilde{I}_{b}]}$ in agreement with idea of "deverse" matrix representation of a bridge channel. The presented approach is independent from any observable and structure of initial states of QR-$A$ and QR-$B$.

Based on discussion above, we decided to use the following optimization objective:
\begin{equation}\label{opt_bridge_obj}
    \vec{\theta}_{*} =  {\rm{argmax}}_{\vec{\theta}}||M_{ab,a'b'}(\vec{\theta})||_{1}; \ \theta_{7}^{2}+\theta_{8}^{2}+\theta_{9}^{2} \gg 0.001.
\end{equation}
To solve the optimization problem in Eqs.(\ref{opt_bridge_obj}), we apply the Adam optimization algorithm \cite{kingma2014adam} with slightly adjusted default parameters.
Solving this optimization, we have found the following optimal parameters for ansatz Eq.(\ref{ANSATZ}) in accordance with Eq.(\ref{opt_bridge_obj}): $\theta_{7} = 0.216  , \theta_{8} = 0.469 , \theta_{9} = 1.023$.
We provide all other values of $\theta_{*}$ and the supporting code in our GitHub repository \cite{github}.

Let us shift to the purification channels.
\begin{figure}[ht!]
\centering
\includegraphics[width= 16 cm]{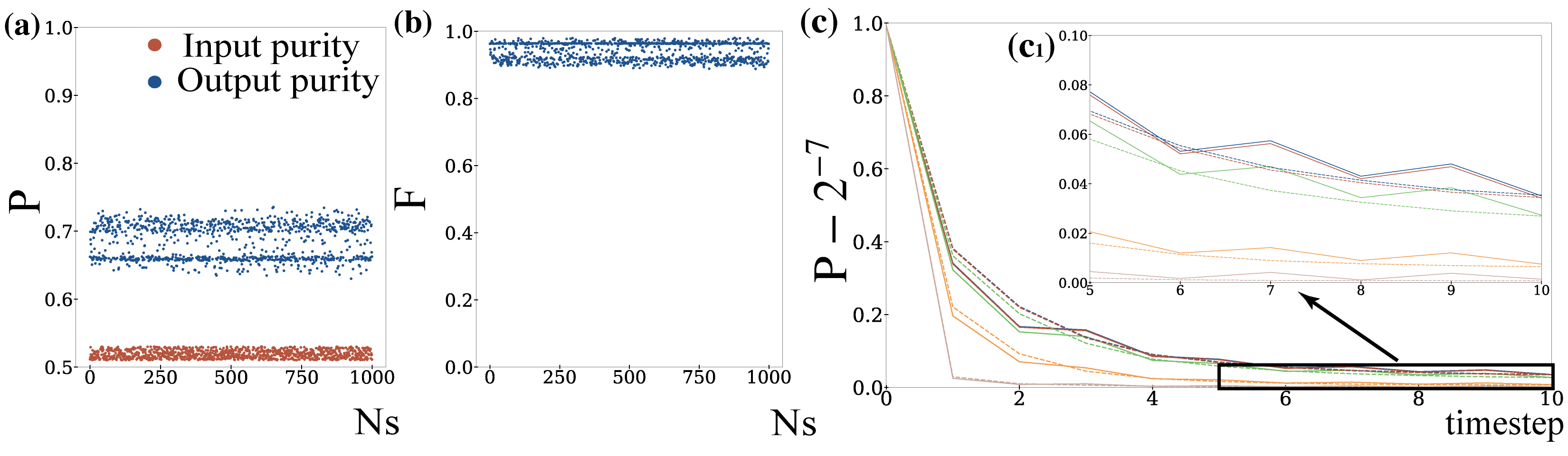}
\caption{Results of numerical optimization for the purification channel $\mathcal{P}$. 
We sampled 1000 qubit states with purity $P\in \{0.51, 0.53\}$ to verify the optimization results. 
In {\textbf{(a)}} we observed a significant increase of $P$ for the corresponding output states, while in {\textbf{(b)}} the fidelity $F$ between the input and output states is close to $0.9$. The panel {\textbf{(c)}} shows the purity for the full QR state $\hat{\varrho}_{AB}$ affected by the depolarization (setup parameters given in Fig.(\ref{ALL_P})). In $\rm{(c_{1})}$ we considered the subsequent 5 timesteps of QRC dynamics with bridge and purification channels as depicted in Fig.(\ref{fig:ESQR}).}
\label{PURITY_ANALYSIS}
\end{figure}
Auxiliary purification channels $\mathcal{P}_{A,B}$ improve the accuracy of the QRC affected by noise. 
Specifically, the purification channels can reduce the number of required measurements (over the QR's ensemble of identical systems) to extract output signals - the average values of some observables $\braket{\hat{O}_{\bullet}}(s)$. 
This follows from the intrinsic structure of the QRC framework.
Indeed, one addresses only to single corresponding qubit QR-"$\bullet$" with the state $\hat{\varrho}_{\bullet} = {\rm{tr}}_{AB/\bullet}(\hat{\varrho}_{AB})$ to perform measurements.
Physically, it means that we have many identical measurement apparatuses that interact with the corresponding QR's sub-ensemble. Alternatively, if we have just one QR system, we can prepare its state using the same sequence of operations multiple times to perform multiple measurement experiments.
Thus, one derives the output signals by estimating $\braket{\hat{O}_{\bullet}}(s) =  {\rm{tr}}_{\bullet}\left(\hat{\varrho}_{\bullet}\hat{O}_{\bullet}\right)$ from statistics of measurements.
It is clear that qubit states $\hat{\varrho}_{\bullet}$ with purity $P = {\rm{tr}}\left(\hat{\varrho}_{\bullet}^{2}\right) \approx 1/2 + \epsilon, \epsilon \ll 1 $ yield $\braket{\hat{O}_{\bullet}}(s) \approx {\rm{const}} + \epsilon \rm{tr}\left(\hat{O}_{\bullet}\ket{\Psi_{\bullet} (s)}\bra{\Psi_{\bullet} (s)}\right)$.
Here the state $\ket{\Psi_{\bullet} (s)}\bra{\Psi_{\bullet} (s)}$ is some pure state ($P_{\ket{\Psi_{\bullet}(s)}} \equiv 1$), and $\rm{const}$ a real number that does not depend on a signal. 
It means that the output signal carrying valuable information has a small magnitude $\sim \epsilon$. Consequently, it requires many measurements to recover a proper output signal. The application of the purification channel seems can help to resolve this issue. In the present work, we assume that we can only purify states of the bridge qubits $a$ and $b$. Without loss of generality, we consider the qubit $a$. 
To optimize the purification channels $\mathcal{P}_{A,B}(\vec{\gamma})$ presented in Eq.(\ref{Purification}) we again address to its matrix representations. 
We use the same parameterization ansatz $\hat{U}(\vec{\gamma})$ - Eq.(\ref{ANSATZ}) as we used before in the case of bridge optimization.
A single qubit state has a well-known geometrical representation on a Bloch sphere \cite{Nielsen2009,Heinosaari2009}: $\hat{\varrho}_{a} =(1/2)\left(\hat{\mathbf{I}} + \sum_{a}r_{a}\hat{\sigma}_{a}\right)$, where $\vec{r}$ is a three dimensional euclidean vector, such that $0 \leq||\vec{r}|| \leq 1$, $||\vec{r}|| = \sqrt{\left(\vec{r},\vec{r}\right)}$. 
Purity of state $\hat{\varrho}_{a}$ is given by $P = (1/2)(1 + ||\vec{r}||^{2})$. 
The action of a purification channel results a transformation of $\vec{r}$ in accordance with:
\begin{eqnarray}\label{purity_transform_for_opt}
\vec{r'} = \vec{q} + \Gamma \vec{r}; \ q_{i} = \frac{1}{2}{\rm{tr}}(\hat{\sigma
}_{i}\mathcal{P}(\vec{\gamma})[{\mathbf{I}}]), \Gamma_{ij} = \frac{1}{2}{\rm{tr}}\left(\hat{\sigma}_{i}\mathcal{P}(\vec{\gamma})[{\hat\sigma}_{j}]\right); \ i,j\in\{1,2,3\}.
\end{eqnarray}
The optimization heuristic is then straightforward. 
We aim to maximally increase the state's Purity and simultaneously preserve as much information as we can. 
As a measure of information "preservation", we use a Fidelity \cite{Nielsen2009, Heinosaari2009}: $F = \rm{tr}(\sqrt{\sqrt{\hat{\varrho}^{'}_{a}}\hat{\varrho}_{a}\sqrt{\hat{\varrho}^{'}_{a}}})$ between qubit's states before and after purification.
Finally, we present an optimization objective $\mathcal{P}(\gamma)$  based on the described geometrical interpretation and Purity - Fidelity trade-off. 
We choose the following form for the optimization objective:
\begin{eqnarray}\label{P_OPT}
\vec{\gamma}_{*} = {\rm{argmin}}_{\gamma}|||\Gamma^{T}(\gamma)\Gamma(\gamma)||_{F}/||\vec{q}_{\gamma}|| - 1|
\end{eqnarray}
The symbol $||\bullet||_{F}$ in Eq.(\ref{P_OPT}) denotes a Frobenius matrix norm. 
Fig.(\ref{PURITY_ANALYSIS}) demonstrates the effectiveness of the proposed numerical optimization.
Note that the trade-off between purity and fidelity in a state purification protocol via sampling methods was explored in work\cite{DiFranco2013}. 
As mentioned, we use the same numerical approach as the one presented in the bridge channel optimization. 
An application of the Adam optimizer yields the following values: $\gamma_{7} = 0.44 , \gamma_{8} = 1.13 , \gamma_{9} = 0.44$. 
Other values of $\vec{\gamma}_{*}$ and all relevant code are provided in ours GitHub repository\cite{github}. 
We utilized the derived parameters $\gamma_{*}$ in the $\textbf{Results}$ to verify usefulness of a purification channel $\mathcal{P}(\gamma_{*})$.

\bibliography{sample}
%For data citations of datasets uploaded to e.g. \emph{figshare}, please use the \verb|howpublished| option in the bib entry to specify the platform and the link, as in the \verb|Hao:gidmaps:2014| example in the sample bibliography file.
\section*{Acknowledgements} 
This work is supported by the Russian Science Foundation under Project No. 20-72-00116. We thank Savva Morozov, Viacheslav Sadykov, Andrei Nomerotski, Andrey Rogachev and Ning Bao for fruitful discussions and support. 

\section*{Author contributions statement}
Both authors contributed equally to the paper.

\section*{Additional information}
\textbf{Competing interests} The authors declare no competing financial interests.

\end{document}